# Potential for Improved Time Resolution Using Very Thin Ultra-Fast Silicon Detectors (UFSDs) [1]

A. Seiden[a*], H. Ren[a], Y. Jin[a], S. Christie[a], Z. Galloway[a], C. Gee[a], C. Labitan[a], M. Lockerby[a],
F. Martinez-McKinney[a], S. M. Mazza[a], R. Padilla[a], H. F.-W. Sadrozinski[a], B. Schumm[a],
M. Wilder[a], W. Wyatt[a], Y. Zhao[a], N. Cartiglia[b]

[a]SCIPP, UC Santa Cruz, Santa Cruz, CA 95064, USA
[b]INFN, Torino, Italia

**Abstract**

Ultra-Fast Silicon Detectors (UFSDs) are n-in-p silicon detectors that implement moderate gain (typically 5 to 25) using a thin highly doped p++ layer between the high resistivity p-bulk and the junction of the sensor. The presence of gain allows excellent time measurement for impinging minimum ionizing charged particles. An important design consideration is the sensor thickness, which has a strong impact on the achievable time resolution. We present the result of measurements for LGADs of thickness between 20 μm and 50 μm. The data are fit to a formula that captures the impact of both electronic jitter and Landau fluctuations on the time resolution. The data illustrate the importance of having a saturated electron drift velocity and a large signal-to-noise in order to achieve good time resolution. Sensors of 20 μm thickness offer the potential of 10 to 15 ps time resolution per measurement, a significant improvement over the value for the 50 μm sensors that have been typically used to date.

## 1. Introduction

Ultra-Fast Silicon Detectors (UFSDs), which allow excellent timing measurements [1], are a recent advance in semiconductor particle detectors. They consist of thin (50 µm being the most common choice) n-in-p Low-gain Avalanche Detectors (LGADs), with detector bias voltage chosen to typically provide internal gain of 5-25 due to a highly doped p++ layer (called the gain layer) between the high resistivity p-bulk and the junction [2, 3], along with broad-band front-end electronics to capture the fast signals. The gain layer has a large electric field coming from a superposition of the field from the full depletion of the p++ implant (a majority of the field) and the voltage on the sensor. This field leads to amplification of electrons only, avoiding an electrical breakdown. The field configuration across the whole sensor has to not only lead to the correct gain in the gain layer but also to a saturated velocity for electrons and a large velocity for holes drifting in the silicon. These provide important constraints on the gain layer doping density. The desired gain layer field depends on the detailed geometry of the gain layer while the field in the bulk should be at least 20 kVolts per cm for a saturated electron drift velocity as well as large hole drift velocity.

---

In this paper, we present measurements of the timing performance of LGADs with thickness of 20 μm, 35 μm or 50 μm. All sensors were fabricated by Hamamatsu Photonics (HPK). Sensors with 50 μm active thickness have by now provided many measurements, including the first measurement of timing performance of 16 ps for a three sensor combination [4]. In the next section of this paper we present some characteristics of the current-source signal for the 50 μm thick detectors. The subsequent section discusses the front-end electronics, which produces the voltage signal that we use to make the timing measurements. The final section presents the results for the three different thickness sensors. Simulation results of the sensor behaviour are derived from [5].

## 2. LGAD Signal

Figure 1 shows a simulated LGAD current pulse for a 50 μm thick sensor and a gain of 20 in the gain layer. The initial very rapid rise is due to the initial motion of the primary electrons and holes produced by the passing minimum ionizing particle. The time of the peak of the signal corresponds to the arrival of the last drifting electron, determined by the detector thickness for a saturated electron drift velocity, which results in a very stable signal shape. This peak time, given by the detector geometry, is independent of the total number of electrons produced or the fluctuations (called Landau fluctuations) in the production process. At the time of the final electron collection, all the holes produced by the gain mechanism are still drifting, since they have a smaller drift velocity than the electrons, providing the first downward slope in the distribution. Once the holes start being collected we get the smooth remaining part of the distribution. We will see that the electronics used for the measurement of signals is not fast enough to capture the very rapidly changing parts of the current signal but rather produces a very smooth rise and then fall for the voltage signal. This is still sufficient to provide very rapid signals that are nearly as fast as the current signal. Figure 2 shows the frequency spectrum of the current signal above 100 Megahertz. The electronics used to measure the signal has a bandwidth of about 2 Gigahertz, which is fast enough to contain most of the signal. The peak height of the current signal is independent of the detector thickness for a fixed gain while the rise-time is proportional to the detector thickness. For the 50 μm thick sensor the simulated 0-100% rise-time is 440 ps.

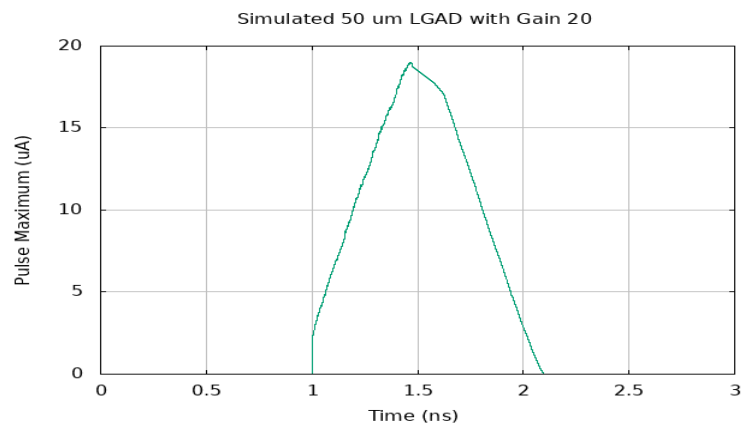

**Fig. 1  Simulated 50 μm thick LGAD current signal.**

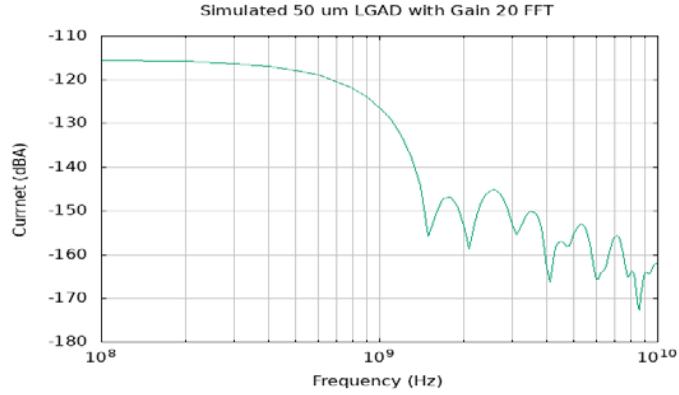

**Fig. 2  Frequency spectrum of the LGAD signal.**

## 3.  Front-end Electronics

The electronics used to measure the signals presented in this paper is a single-channel trans-impedance amplifier of 22 ohm input impedance. It is made of discrete components and is meant to be wire bonded to the LGAD pad. A schematic is shown in Figure 3.

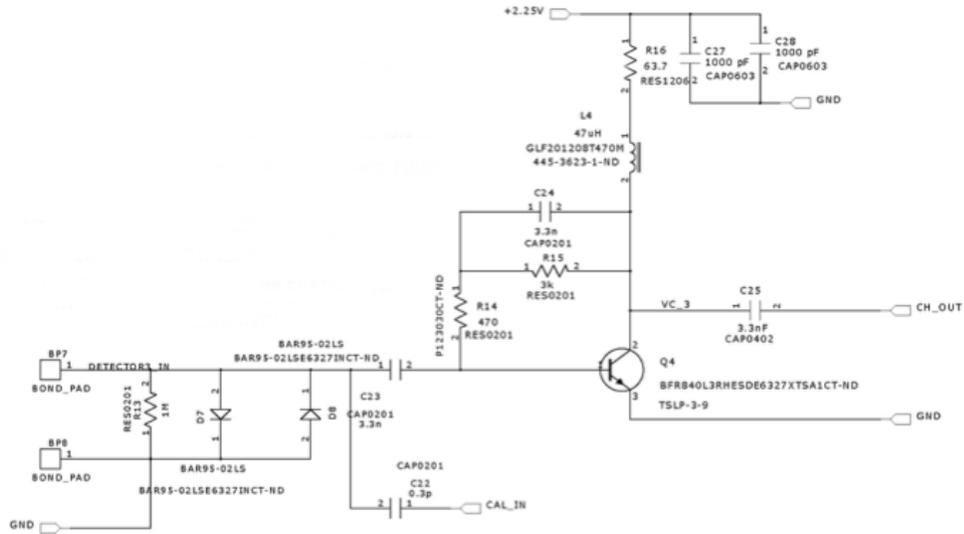

**Fig. 3  Schematic of single-channel front-end amplifier.**

The amplifier noise is completely dominated by the Johnson noise of the resistive components and given the fixed bandwidth of the amplifier the noise is independent of the detector capacitance. Figures 4 (pulse shape) and Figure 5 (rise-time for 10-90% rise and 0-100% rise) show the simulated response of the amplifier to the LGAD pulse in Figure 1 as a function of the detector capacitance.

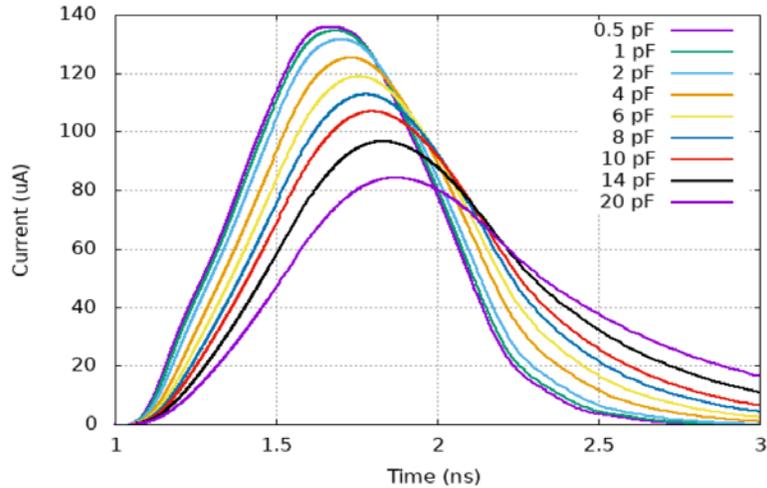

**Fig. 4  LGAD pulse out of amplifier for different sensor capacitance.**

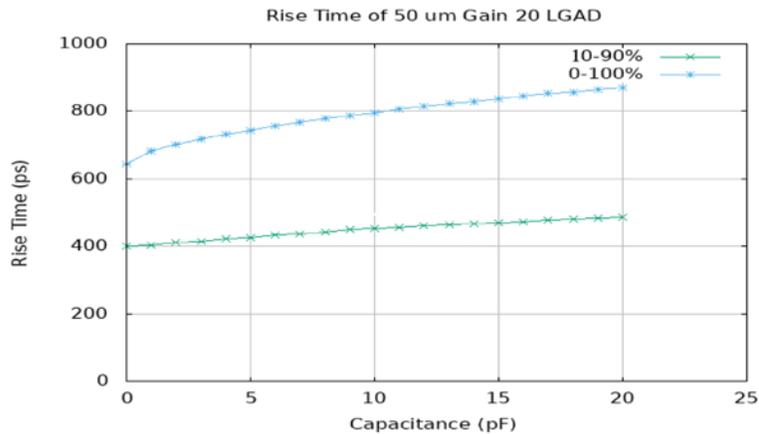

**Fig. 5  Rise-time (both 10-90% and 0-100%) for signals in Fig. 4.**

From Figures 4 and 5 we see that the output of the amplifier is nearly independent of the detector capacitance for values up to a few pF. It is therefore very useful for comparing different sensor types in this capacitance range with only small changes in signal shape due to the detector capacitance. All the sensors discussed later have a capacitance < 5 pF.

Figure 6 shows LGAD pulses from a 50 μm thick sensor using the single-channel amplifier. The pulses, 100 in total, are scaled to all have the same maximum height and show how uniform the pulse shape is. The data were collected using a $^{90}$Sr beta-source with a trigger provided by a fast HPK LGAD. We use this setup for data collection for results presented in the next section of the paper and call it the "beta beam", described in more detail in [6]. The figure illustrates a number of properties of LGAD pulses. The time of the signal maximum depends only on the detector thickness and is independent of the gain and pulse height as long as the electron drift velocity is saturated. The pulse height maximum, not shown in the figure, is proportional to the gain. The rising part of the pulse, 10-90% rise, is to good approximation a ramp with equation: $V(t) = t\,(P_{MAX}/T_{MAX})$. From the width of the baseline in Figure 6 it is also clear that the LGAD provides an excellent signal-to-noise ratio.

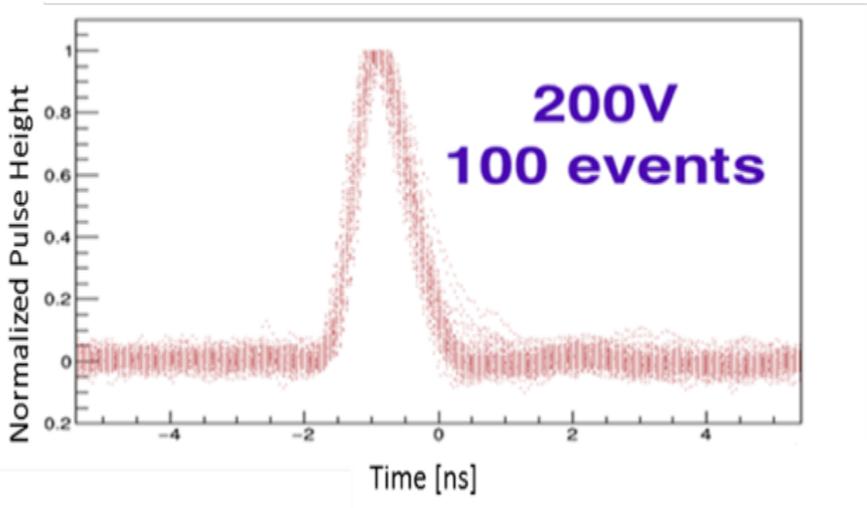

**Fig. 6 Total of 100 LGAD pulses overlaid, all with peak normalized to 1.**

Typically the voltage pulse is discriminated to establish the time of passage of the incident particle. This is done by imposing a threshold, $V_{TH}$, with the time given by the value of t satisfying $V(t) = V_{TH}$. For an LGAD the time resolution is limited mainly by noise in the electronics (called jitter) and Landau fluctuations in the charge deposition within the sensor. We quantify the electronics noise by a standard deviation σ. If we imagine the noise changing the threshold from pulse to pulse, with $\delta V_{TH} = \sigma$, we can get a simple relation for the jitter contribution by differentiating the threshold equation, giving: $\sigma = (dV/dt)\,\delta t$, where dV/dt is evaluated at the time where $V(t) = V_{TH}$ and δt is the resulting time resolution. Therefore: $\delta t = \sigma/(dV/dt)$. We have not used here any characteristics of a specific amplifier and this result is quite general [7]. The resolution will therefore in the general case depend on the choice of threshold and amplifier. In applications we expect that $V_{TH}$ will be chosen smaller than about half the most likely value of the signal peak height to maintain high efficiency given the fluctuations in pulse height and larger than a value which is determined by the electronics noise. A threshold choice, $V_{TH,}$ at the time that maximizes dV/dt should give the best result for the jitter.

For the ramp pulse approximation to the voltage from the single-channel amplifier, we calculate dV/dt, which is $P_{MAX}/T_{MAX}$, independent of the choice of threshold in this specific case. Therefore the expected jitter contribution to the resolution is: $\delta t = \sigma/(P_{MAX}/T_{MAX})$. This can be conveniently written as: $\delta t = T_{MAX}/(P_{MAX}/\sigma)$. Defining $(P_{MAX}/\sigma)$ as the signal-to-noise ratio (SNR) gives: $\delta t = T_{MAX}/SNR$. Since $T_{MAX}$ is proportional to the sensor thickness we get that the jitter is proportional to the sensor thickness; also the SNR is proportional to the sensor gain so a larger gain gives a smaller jitter. In addition the jitter is proportional to the electronics noise. In analysing signals we have usually established the threshold crossing time using a constant fraction algorithm applied to pulses saved using a storage scope. If we write the time of the pulse versus the signal amplitude as $t = g(V)$, a dependence on the pulse shape of the form $g(V/P_{MAX})$ for the rising part of the pulse as in Figure 6 leads to a fixed time resolution by choosing a constant fraction $f = V/P_{MAX}$. This does not depend on the form of the function g and leads to a choice of time $t = g(f)$. We find that choosing the constant fraction anywhere between 30% and 70% gives close to the same resolution. This is expected given that $\delta t$ doesn't depend on the threshold choice for the single-channel amplifier when we approximate the pulse as a ramp. The constant fraction choice and the resulting impact on the time resolution for 50 μm thick sensors is explored in [6] where both data and simulations using [5] are presented. For constant fraction values below 30%, dV/dt decreases noticeably and it decreases rapidly below 10%. Despite this decrease, for large gain, for example larger than 30, choosing a very small constant fraction, for example 10% for a 50 μm thick sensor, results in better time resolution since this is essentially a way to minimize the effect of Landau fluctuations, discussed below. For large gain, the jitter contribution is very small and the Landau fluctuations determine the resolution.

Besides jitter, the limitation to the resolution is provided by the Landau fluctuations. These have been evaluated using a Monte Carlo program [5]. The result is that the contribution of these to the resolution is proportional to the sensor thickness for thin sensors for a constant fraction choice larger than about 20%. Given that $T_{MAX}$ is also proportional to the detector thickness we can write the Landau contribution as $\alpha_L T_{MAX}$. Adding the two contributions to the resolution, jitter and Landau fluctuations, in quadrature we get finally a prediction for the resolution for thin detectors as: $\sigma_t = T_{MAX}[\sqrt{((1/SNR^2) + \alpha_L^2)}]$. We compare this formula to our data in the next section. SNR and $\sigma_t$ are directly measured using beta-beam data while $T_{MAX}$ and $\alpha_L$ are fit to the data individually for each sensor type. Fitting these two parameters allows the best description of the data using the simple resolution formula. We expect that $T_{MAX}$ should be ~ 10-90% pulse rise-time of the given sensor signals and that the resolution flattens out at very large SNR at a value $\alpha_L T_{MAX}$, which is ~ 30 psec for a 50 μm thick sensor based on previous data [4, 6]. We expect $\alpha_L$ to be approximately a constant independent of sensor thickness as long as the electron drift velocity is saturated so that the total drift time is proportional to the sensor thickness.

## 4. Sensor Measurements

We have performed a large number of measurements on LGADs of various thicknesses using the beta-beam setup to quantify the timing performance. The detector voltage was varied to provide various gain values resulting in corresponding values of the SNR defined as the signal peak divided by the noise. Since the amplifier noise is approximately constant up to reasonably large gain values, the value of the SNR depends primarily on the magnitude of the peak of the signal. A gain of about 12 results in a SNR of 20, with an uncertainty in the gain determination of about 20%. The SNR is determined directly from the measured pulse height and noise and does not involve integrating over the full pulse and comparison to sensors with no gain, as required for the gain determination. The LGADs used in this study were circular pads of thickness 20 μm or 35 μm and square pads with a thickness of 50 μm. The 50 μm thick sensors come in two varieties (called 3.1 and 3.2), which differ from each other in the doping profile and amount of dopant in the gain layer but are expected to give similar results for the same SNR as long as the electron drift velocity is saturated. The achievable gain for the 20 μm thick sensor was not as large as for the others as the gain layer doping was not optimal.

Table 1 has a description of the sensors used. This includes the sensor name, thickness, dimensions of the pad used to collect the charge, capacitance of the sensor, and the measured 10-90% rise-time of the signal for a saturated electron drift velocity. For the data analysed below, the rise-time values have a standard deviation of between 6% and 8% for the various sensor types. We use a cut, which is a box of full-width = 80 ps for the 20 μm thick sensor and 100 ps for the others, centered on the average rise-time, to select data with a saturated electron drift velocity. The average rise-time values agree to 10% with the expectation that they are proportional to the sensor thickness with a small dependence on the sensor capacitance. For example, the different rise-times for the two types of 35 μm thick sensors are due to the differing capacitance values as the sensor with the larger capacitance can be expected to have a larger rise-time by ~ 10% according to Figure 5. The range of rise-times from 178 picoseconds for the thinnest sensors to around 400 picoseconds for the 50 μm thick sensors allows a good exploration of the impact of sensor thickness and rise-time on the timing performance.

Our analysis always uses a 50% value for the constant fraction determination. The trigger is provided by a sensor with 17 ps resolution, which is subtracted in quadrature from the measured resolution to arrive at the timing resolution $\sigma_t$. Figure 7 shows a typical pulse for a 20 μm LGAD at a representative voltage. Pulse shapes under a range of conditions for 35 μm and 50 μm thick sensors can be found in [6, 8].

**Table 1: Description of Sensors Used in Study**

| Sensor Name | Thickness | Pad Dimensions | Capacitance | Signal: 10-90% rise - time |
|---|---|---|---|---|
| H20 | 20 microns | Circle of radius = 0.37 mm | 2.6 pF | 178 psec. |
| G35 | 35 microns | Circle of radius = 0.37 mm | 1.5 pF | 262 psec. |
| B35 | 35 microns | Circle of radius = 0.63 mm | 4.4 pF | 306 psec. |
| 3.1 | 50 microns | Square 1.3mmx1.3mm | 3.9 pF | 398 psec. |
| 3.2 | 50 microns | Square 1.3mmx1.3mm | 3.9 pF | 404 psec. |

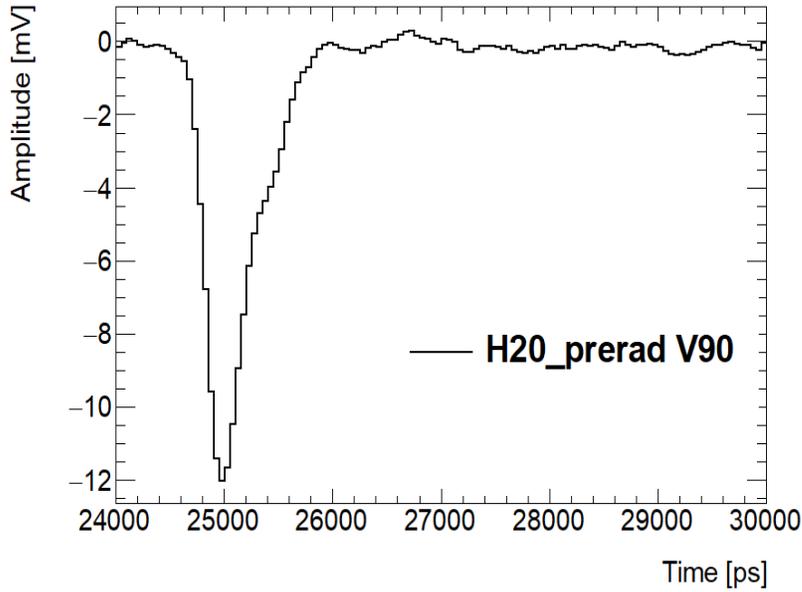

Fig. 7  Pulse for 20 μm thick LGAD. Gain = 8.

Figures 8, 9, 10, respectively, show the time resolution achieved versus SNR for the five sensor types. Also shown is a fit to the data of the form $\sigma_t = T_{MAX}[\sqrt{(1/SNR^2) + \alpha_L^2}]$. The data come from sensors with varying conditions, including irradiated sensors (up to $2.5 \times 10^{15}$ neq/cm$^2$ radiation levels) and runs at various temperatures (from -20 degrees to -30 degrees). The data for each sensor type have similar rise-time, key to determining the resolution and it is not necessary to distinguish the data runs based on irradiation conditions, an interesting result of the analysis. Runs where the noise was very large for voltages near breakdown have been excluded.

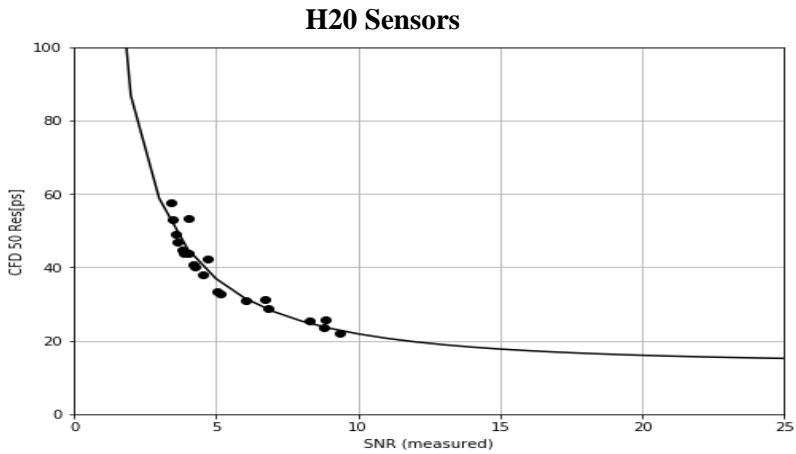

Fig. 8  Time resolution versus SNR for 20 μm thick LGAD. Gain > 2.5 was required.

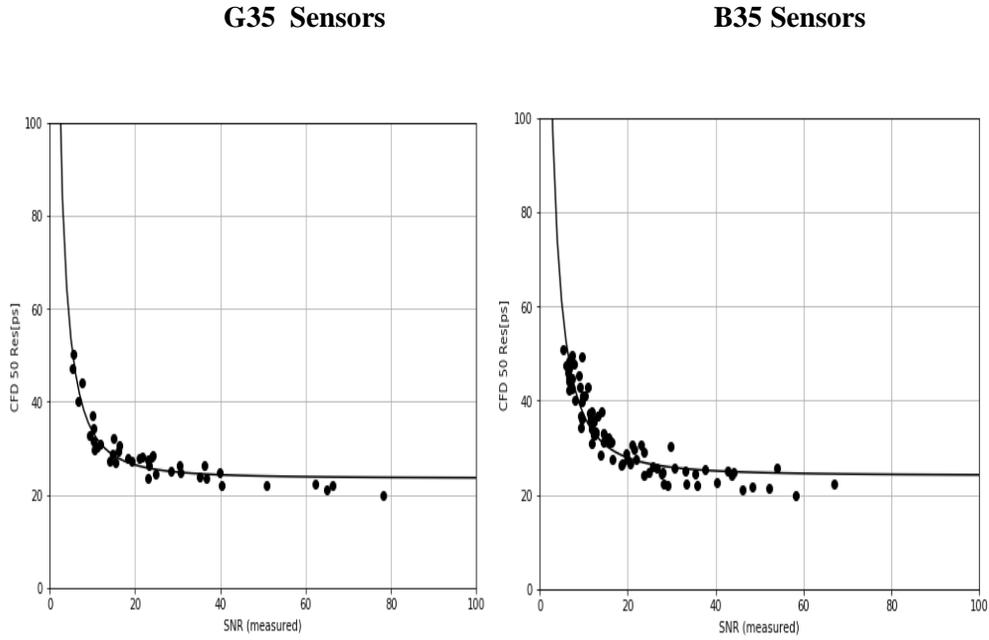

**Fig. 9** Time resolution versus SNR for 35 μm thick LGAD, with G35 to the left and B35 to the right. Gain > 4 was required.

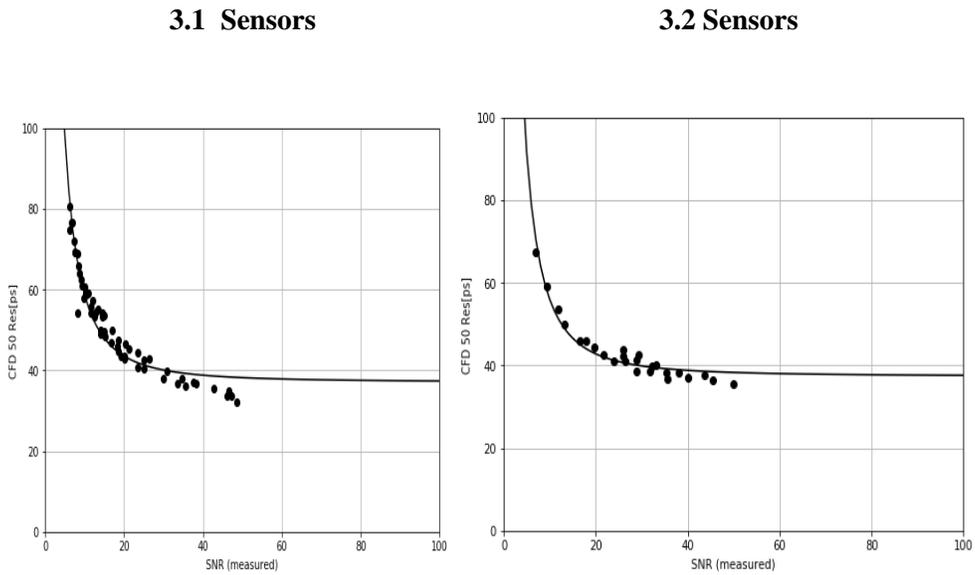

**Fig. 10** Time resolution versus SNR for 50 μm thick LGAD, with 3.1 to the left and 3.2 to the right. Gain > 4 was required.

Table 2 lists the results of the fits for the five sensor types. The table lists the fit values for $T_{MAX}$ and $\alpha_L$ as well as the product $\alpha_L T_{MAX}$, which is the Landau floor. The errors on the two fit parameters are based on a choice of 3 ps error per measured point, which yields a chi-square per degree-of-freedom varying from 0.34 to 1.2 depending on the sensor type. The 3 ps error is based on the variation in resolution when we re-measure points and the variation in rise-times in the data for a given sensor type. The fits indicate that $\alpha_L$ is approximately a constant, equal to 0.085, across sensor types. However, looking at the data, the fits are typically a few picoseconds too large in the region of the Landau floor when the SNR ratio is larger than about 40.

**Table 2: Fit Values for Parameters Determining the Timing Resolution**

| Sensor | $T_{MAX}$ | $\alpha_L$ | $\alpha_L T_{MAX}$ | Uncertainty on $T_{MAX}$ | Uncertainty on $\alpha_L$ |
|---|---|---|---|---|---|
| H20 | 172 psec. | 0.079 | 13.5 psec. | 6 psec. | 0.019 |
| G35 | 240 psec. | 0.098 | 23.5 psec. | 10 psec | 0.006 |
| B35 | 281 psec. | 0.086 | 24.1 psec. | 6 psec | 0.004 |
| 3.1 | 456 psec | 0.081 | 37.1 psec | 9 psec. | 0.003 |
| 3.2 | 417 psec. | 0.090 | 37.4 psec. | 12 psec. | 0.003 |

The data indicate that the signal-to-noise is the primary parameter determining the time resolution once the signal rise-time is fixed, with a floor determined by the Landau fluctuations. The different thickness sensors, have very different time resolution for a given signal-to-noise, tracking the different $T_{MAX}$ values versus thickness. The factor multiplying $T_{MAX}$, sqrt($1/SNR^2 + \alpha_L^2$), has two dimensionless terms and results in the approach to the Landau floor being independent of detector thickness. For example a SNR of 20 gives a resolution which is 16% higher than the floor for $\alpha_L = 0.085$. A SNR ratio = 37 is required to have a resolution 5% larger than the floor. In terms of gain, a gain of 20, which yields a SNR of about 37 for our data, is required to arrive at a time resolution within about 5-10% of the Landau floor. The value of the Landau floor is proportional to the sensor thickness providing a strong motivation for the use of thin sensors.

For the 50 μm thick sensor type 3.1 we have a large amount of data, including some data with the velocity not saturated. The data in Figure 10 have a requirement that the 10-90% rise-time lie between 350 and 450 ps. We can also look at data where this rise-time is between 450 and 550 ps. Figure 11 shows the data for sensor type 3.1 with the two separate rise-time selections as well as the data for sensor type 3.2 with the standard rise-time selection of 350-450 ps. The results for 3.1 and 3.2 are close to each other as expected and the data for 3.1 with the larger rise-time has poorer resolution for SNR ~ 10 to 20. The Landau floor for all measurements is however the same, determined by the sensor thickness.

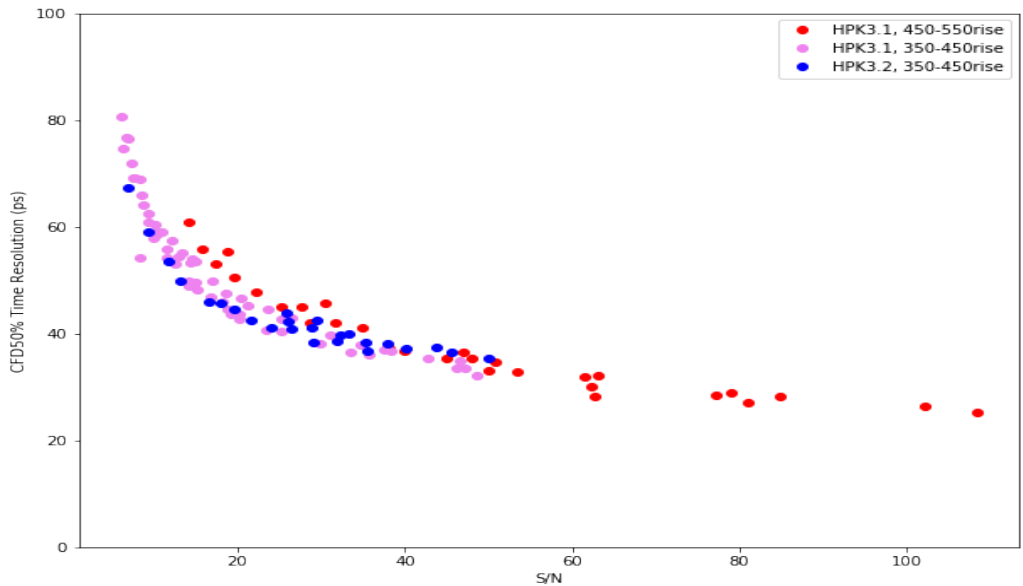

**Fig 11 Time resolution versus signal-to-noise ratio for sensor type 3.1 with selection on the 10-90% rise-time of 350-450 ps (in violet) and separately 450-550 ps (in red). Included also are data for sensor type 3.2 with rise-time between 350-450 ps (in blue).**

## 5. Conclusions

We have presented data collected using LGAD sensors of varying thickness, all read-out through the single-channel amplifier in the beta-beam setup. The time resolution that can be achieved is dependent on the electronics jitter and the Landau fluctuations in the sensor. The jitter can be calculated in terms of the amplifier noise $\sigma$ and slew rate of the voltage signal as $\delta t = \sigma/(dV/dt)$. The Landau fluctuations are proportional to the sensor thickness for thin sensors and provide a contribution to the time resolution in quadrature with the jitter. For the single-channel amplifier, where the voltage ramp is given to good approximation by $V(t) = t\,(P_{MAX}/T_{MAX})$, the two contributions can be combined to give $\sigma_t = T_{MAX}\,[\sqrt{(1/SNR^2) + \alpha_L^2}\,]$. The product $\alpha_L T_{MAX}$ is the Landau floor for a given sensor thickness. The constant $\alpha_L$ has been determined to be about 0.085.

Given that $T_{MAX}$ is proportional to the sensor thickness, thin sensors offer improvement in time resolution for both terms determining the resolution. The measured Landau floor for 35 μm thick sensors is about 23 ps. Using the fit, the predicted Landau floor for the 20 μm thick sensor is about 13 ps. A smaller pad for a 20 μm sensor, with smaller capacitance, should allow a somewhat reduced rise-time and a smaller resolution value. Combining this with small adjustments in thickness and an optimized doping layer concentration should allow a resolution close to 10 ps. A major challenge for getting the best time resolution from thin sensors is the need for high bandwidth in the electronics, given the fast current signal, coupled with small noise to achieve a large SNR. Optimizing the electronics is a complementary challenge to producing an optimized sensor.

**Acknowledgements**

This work was supported by the United States Department of Energy, grant DE-FG02 04ER41286.

Part of this work has been financed by the European Union's Horizon 2020 Research and Innovation funding program, under Grant Agreement no. 654168 (AIDA-2020) and Grant Agreement no. 669529 (ERC UFSD669529), and by the Italian Ministero degli Affari Esteri and INFN Gruppo V.